\documentclass[lettersize,journal]{IEEEtran}
\usepackage{amsmath,amsfonts,amssymb}
\usepackage{graphicx}
\usepackage{subfig}
\usepackage{algorithmicx}
\usepackage{algorithm}
\usepackage{algpseudocode}
\usepackage{textcomp}
\usepackage{makecell}
\usepackage{multirow,multicol}
\usepackage[colorlinks]{hyperref}
\usepackage[table]{xcolor}
\usepackage{threeparttable}

\begin{document}

\title{On Fundamental Limits for Fluid Antenna-assisted Integrated Sensing and Communications for Unsourced Random Access}

\author{Zhentian Zhang,~\IEEEmembership{Student Member,~IEEE}, 
            Kai-Kit Wong,~\IEEEmembership{Fellow,~IEEE}, 
            Jian Dang,~\IEEEmembership{Senior Member,~IEEE}, \\
            Zaichen Zhang,~\IEEEmembership{Senior Member,~IEEE}, and 
            Chan-Byoung Chae,~\IEEEmembership{Fellow,~IEEE}
\vspace{-5mm}

\thanks{ }
\thanks{Zhentian Zhang, Jian Dang, Zaichen Zhang are with the National Mobile Communications Research Laboratory, Frontiers Science Center for Mobile Information Communication and Security, Southeast University, Nanjing, 210096, China. Jian Dang, Zaichen Zhang are also with the Purple Mountain Laboratories, Nanjing 211111, China (e-mail: \{zhangzhentian, dangjian, zczhang\}@seu.edu.cn).}
\thanks{Kai-Kit Wong is affiliated with the Department of Electronic and Electrical Engineering, University College London, Torrington Place, WC1E 7JE, United Kingdom and also affiliated with Yonsei Frontier Lab, Yonsei University, Seoul, Korea (e-mail: kai-kit.wong@ucl.ac.uk).}
\thanks{Chan-Byoung Chae is affiliated with Yonsei Frontier Lab, Yonsei University, Seoul, Korea (e-mail: cbchae@yonsei.ac.kr).}
\thanks{Corresponding authors: Jian Dang (dangjian@seu.edu.cn), Zaichen Zhang (zczhang@seu.edu.cn).}}
%
%

\maketitle
\begin{abstract}
This paper investigates the unsourced random access (URA) problem for integrated sensing and commutations (ISAC). Recent results reveal that conventional multiple access strategies for ISAC such as treating interference as noise (TIN) and time-division multiple access (TDMA) can be easily overwhelmed and fail to support the increasingly surging number of active users. Hence, the unsourced ISAC (UNISAC) system model has emerged as a promising enabler for the future ISAC networks. To advance this work, we adopt a more realistic channel model and propose to utilize fluid antenna system (FAS) for UNISAC. The achievable performance bound and floor of the proposed FAS-UNISAC are derived to validate the great potential. Our results demonstrate that promising improvement on the available user volume and the sensing and communication capability can be obtained due to the spatial diversities inherent within fluid antenna. 
\end{abstract}

\begin{IEEEkeywords}
Fluid antenna system (FAS), integrated sensing and commuication (ISAC), unsourced ISAC (UNISAC), unsourced random access (URA). 
\end{IEEEkeywords}

\section{Introduction}
\subsection{Background}
Massive machine-type communications (mMTC) \cite{mMTC_6G1,mMTC_6G2} is one of several key use cases in the upcoming sixth generation (6G) communication networks. The surge in the number of devices and their variety in processing power make it difficult for coordination-based multiple access techniques to work \cite{URA_China_Comm}. Coordination-oriented system designs are destined to fail for massive access. Specifically, the information bit per channel use for each user approaches zero with the number of devices growing to infinity \cite{WYP_Survey}. Alternative multiple access techniques are required if the demands of future networks are to be met.

To tackle this issue, the concept of unsourced/uncoordinated random access (URA) \cite{Polyanskiy} has emerged to be able to simultaneously support many devices without much preprocessing of their signals. In URA, all users share a common codebook for information projection. The signals are then directly transmitted without any coordination. The central receiver decodes a list of messages from the received signal based on a common codebook without user identification. The system performance is measured by the per-user-probability of error (PUPE), i.e., the number of erroneous detection on codewords or the frame error rate. Under the constraint of finite blocklength, the goal is to reach the prescribed PUPE target with energy-per-bit as low as possible. Thereafter, the achievability bound under various URA channel models are discussed \cite{Bound_1,Bound_2,Bound_4,Bound_5}.

\subsection{URA and Integrated Sensing and Communication (ISAC)}
The discussion of 6G also observes the convergence of communication and radar systems \cite{ISAC1}, leading to the paradigm of integrated sensing and communication (ISAC) \cite{ISAC,ISAC2,ISAC3}. These developments have led to the work in \cite{ISAC-URA} introducing the novel unsourced ISAC (UNISAC) model in which some performance bound was derived as benchmarks for practical protocol designs. In UNISAC, an abundant number of communication users (CUs) and sensing users (SUs) access the system through uncoordinated uplink transmission.\footnote{While active sensing is assumed in the UNISAC model \cite{ISAC-URA}, i.e., users transmit codewords actively instead of being detected passively from the radar echoes, the model is still reasonable when there are simultaneous echoe signals from a large number of users.} The key is to understand the minimum required energy-per-user to reach the prescribed sensing and communication system metrics. 

Compared with the coordination-based ISAC system design, based on conventional multiple access schemes such as time-division multiple access (TDMA), or treating interference as noise (TIN), UNISAC thrives when the numbers of CUs and SUs increase, and in this situation, existing coordination-based protocols fail, either with the required energy level becoming prohibitively high or the system being overwhelmed by the large signal flows unable to conduct decoding and detection. Comparatively, UNISAC offers a more realistic multiple access infrastructure for ISAC with massive connectivity.

\subsection{Challenges}
In UNISAC, of particular importance is the finite coherence blocklength which represents the duration where the channel coefficients remain unchanged \cite{Blocklength1,Blocklength2}. Typically, the coherence time approximately equals to $1/4D_s$ where $D_s$ is the maximal Doppler spread. For a $2~{\rm GHz}$ carrier, the coherence time ranges from $1~{\rm ms}$ to $45~{\rm ms}$ corresponding to the mobility speed between $3~{\rm km/h}$ and $120~{\rm km/h}$. On the other hand, the sampling frequency normally ranges between $100~{\rm kHz}$ and $500~{\rm kHz}$ in outdoor environments, which represents a finite blocklength of $100$ to $20,000$ channel uses. For UNISAC, the expectation is that $100$ bits are delivered over $5000$ channel uses, giving rise to a spectral efficiency of $0.02$ bits/channel-use. This clearly does not indicate a great throughput but the key here is to allow an enormous number of devices to share the same physical channel and deliver the same.

There are limitations on the existing theoretical results of UNISAC. First, the existing achievability bound is built on the channel model with only a line-of-sight (LOS) component for both communication and sensing. Though it is true that such model is commonly assumed for sensing using radar signals, it is well known that the channel condition can be more complex for communications, and channel models consisting of both LOS and non-LOS (NLOS) components are more appropriate. Importantly, the degree-of-freedom (DOF) achievable by this channel mixture needs to be explored. However, the channel model with LOS and NLOS mixtures is hard to analyze. 

Additionally, the sensing estimation model is normally built on the assumption of having a uniform linear array (ULA) for reception. To minimize channel correlation between the array elements, a half-wavelength spacing between the elements is normally assumed. As a result, the physical size of ULA will be increased if more array elements are deployed to enhance the estimation accuracy, which is impractical. How to achieve better sensing performance with a fixed physical antenna size remains an active research endeavour.

\subsection{UNISAC aided by Fluid Antennas}
To fundamentally improve the performance of UNISAC, in this paper, we consider the use of fluid antennas at the central receiver. Fluid antenna system (FAS) represents any software-controllable fluidic, conductive or dielectric structure that has the ability to change its radiation characteristics according to its needs \cite{FAS_CL_2,FAS1}. Focusing on position reconfigurability in antenna, FAS was first introduced to wireless communications in \cite{Wong-2020cl} and \cite{FAS}. Since then, lots of efforts have been made to understand the performance limits of FAS-assisted channels, see e.g., \cite{Psomas-dec2023,Khammassi-2023,Vega-2023,Ye-2023,New-twc2023,Ghadi-2024}. FAS has also found applications in rethinking multiple access \cite{FAS4,FAS2,FAS3}. In recent results \cite{Wang-fasisac2023,FAS5,FAS6,Zou-2024,Meng-2025}, FAS has also been applied for ISAC. Channel estimation for FAS is another key topic, which has led to the methods in \cite{Hao-2024,Dai-2023}.

It would be interesting to see how much FAS can improve UNISAC over using ULA. This motivates us to propose and investigate the FAS-based UNISAC (FAS-UNISAC) paradigm. In this model, multiple antenna array elements or ports can be set without abiding by the half-wavelength rule to achieve better performance. Specifically, the proposed FAS-UNISAC network model is aimed to tackle the challenges with respect to massive connectivity without coordination among users. But the optimization of FAS for UNISAC is not straightforward. Different optimization models on communication and sensing are constructed with analytical solutions. The main contributions of this work are summarized as follows:
\begin{itemize}
\item We present propositions on the achievable performance bound and performance floor for FAS-UNISAC. In our results, a practical channel model with LOS and NLOS components is considered. Meanwhile, a universal upper bound for FAS-based sensing is also derived.
\item For communication task, the relationship between the FAS channel gain and the detection error probability is investigated. The results indicate that the joint detection errors on communication and sensing are inversely proportional to the channel gain brought by FAS.
\item For sensing task, a novel estimation model is established exploiting the covariance information embedded in sensing signals. The inherent spatial diversity in fluid antenna enables a wider receiving aperture by exploiting a virtual array of coherent signals in the covariance domain.
\end{itemize}

The remainder of this paper is organized as follows. Section \ref{sec:model} describes the FAS-UNISAC system model with systematic metrics. Section \ref{sec:pro} presents the achievable bound with essential definitions. In Section \ref{sec:com}, the optimization and analytical solution on the communication tasks for FAS-UNISAC are provided. Then in Section \ref{sec:Sen}, the sensing model along with its optimization and analytical solutions are given. Numerical results with benchmarks are illustrated and discussed in Section \ref{sec:sim}. Finally, conclusions are drawn in Section \ref{sec:conc}.

\textit{Notations:} Lower-case and upper-case bold letters denote column-wise vector $\boldsymbol{a}$ and matrix $\boldsymbol{A}$. The element at the $m$-th row and the $n$-th column is denoted by $[\boldsymbol{A}]_{m,n}$. The transpose and Hermitian of matrix $\boldsymbol{A}$ are denoted by $\boldsymbol{A}^{\mathrm{T}}$ and $\boldsymbol{A}^{\mathrm{H}}$. $\boldsymbol{U}_M$ is an $M\times M$ identity matrix. For a complex scalar $s$, its modulus is given by $|s|$ and its real part is given by $\mathrm{Re}(s)$. The $l_0$-norm, $l_1$-norm and $l_\infty$-norm of matrix $\boldsymbol{A}$ are denoted as $\|\boldsymbol{A}\|_0$, $\|\boldsymbol{A}\|_1$, $\|\boldsymbol{A}\|_\infty$. $F_{\chi^2}(t)$ and $F_{\chi^2}^{-1}(t)$ denote cumulative distribution function of chi-squared distribution with $t$ degrees of freedom and its inverse respectively. $\chi_T^2$ denotes a random variable with chi-square distribution by $T$ degrees of freedom. Function $(\cdot)^{\mathrm{R}}$ reverses the order of elements in a vector.

\section{FAS-UNISAC Model Descriptions}\label{sec:model}
\begin{figure}[htp]
\centering
\includegraphics[width=3.5in]{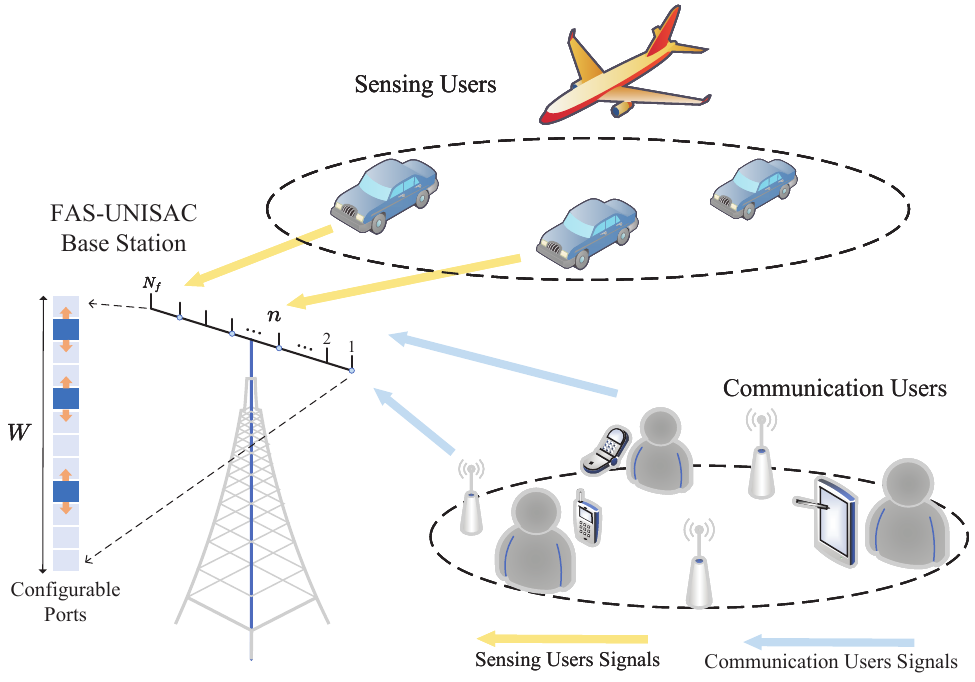}
\caption{An illustration of the FAS-UNISAC network.}\label{fig:system}
\end{figure}

Following the train-of-thought in \cite{ISAC-URA}, fundamental limits of FAS-UNISAC under more practical channel models (including LOS and NLOS propagation components) are introduced without limiting the content to any specific techniques. A possible system model is depicted in Fig.~\ref{fig:system}. We first define the essential parameters in a generalized model and then present the proposition for the achievable results.

For UNISAC, a common codebook $\boldsymbol{A}\in \mathbb{C}^{ 2^{A_{c}+A_{s} \times L} }$ with $2^{A_{c}+A_{s}}$ rows is shared. CUs select codewords with $A_{c}$ bits from the first $2^{A_{c}}$ rows and SUs select codewords from the last $2^{A_{s}}$ rows with $A_{s}$ bits. Neglecting error from asynchronous transmission, the received signal $\boldsymbol{Y}\in \mathbb{C}^{M\times L}$ is written as
\begin{equation}\label{eq:6-1}
\boldsymbol{Y}=\sum_{j\in \{\mathcal{A}_c,\mathcal{A}_s\}}\boldsymbol{g}_{j}\boldsymbol{a}_{j}+\boldsymbol{Z},
\end{equation}
where $\boldsymbol{a}_j\in \mathbb{C}^{1\times L}$ denotes the $j$-th row of $\boldsymbol{A}$ and $\mathcal{A}_c \subset \{1,2,\ldots,2^{A_{c}}\}$ and $\mathcal{A}_s \subset \{1,2,\ldots,2^{A_{s}}\}$, $\boldsymbol{Z}$ is the additive white Gaussian noise (AWGN) with element-wise distribution of $\mathcal{CN}(0,\sigma^2)$. The averaged power per-channel use of CUs and SUs is denoted by $\bar{P}_c$ and $\bar{P}_s$. The fluid antenna at the receiver consists of $N_f$ reconfigurable ports. The channel responses with LOS/NLOS components \cite{FAS,FAS_CL_2} are
\begin{equation}\label{eq:6-2}
g_{j,n}=\underbrace{\sigma_{j,0}e^{-j\frac{2\pi(n-1)W}{N_f-1}\cos\theta_{j,0}}}_{\text{LOS component}}+\underbrace{\sum_{l=1}^{L_s}\sigma_{j,l}e^{-j\frac{2\pi(n-1)W}{N_f-1}\cos\theta_{j,l}}}_{\text{NLOS component}},
\end{equation}
in which there are one LOS path and $L_s$ scattering or NLOS paths. For LOS, $\sigma_{j,0}=\sqrt{\frac{K\Omega}{K+1}}e^{i\alpha_j}$, $K$ is the Rice factor, $\alpha_j$ denotes the arbitrary phase of the LOS path, and $\theta_{j,0}$ represents the azimuth angle-of-arrival (AOA) of the specular component of the $k$-th device. For the scatterers, $\sigma_{j,l}$ is the complex coefficient of the $l$-th path of the $j$-th user, $\theta_{j,l},l\in[1:L_s]$ represents the azimuth AOAs of the $l$-th path of the $j$-th device, and $\sigma_{j,l}$ is the complex coefficient of the $l$-th path of the $j$-th device satisfying $\sum_{l=1}^{L_s}|\sigma_{j,l}|^2=\frac{\Omega}{K+1}$. Constant $\Omega$ denotes the channel strength and is fixed to $1$ without loss of generality. 

For UNISAC, the central receiver, or the base station (BS), aims to restore the binary messages of CUs and estimate the AOA of SUs. Aligning the paradigm in \cite{ISAC-URA}, the system performance is evaluated in terms of PUPE, $\epsilon$, and the mean-square error of AOA (MSEAOA), defined as
\begin{equation}\label{eq:1}
\begin{aligned}
\mathrm{PUPE} &=\frac{\mathbb{E}\{\mathcal{L}_{c,e}+\mathcal{L}_{s,e}\}}{K_c+K_s}, \\
\mathrm{MSEAOA} &=\frac{1}{|\mathcal{L}_{s,d}|}\sum_{\theta_k\in \mathcal{L}_{s,d}}\mathbb{E}\{|\cos\theta_k-\cos\hat{\theta}_i|^2\},
\end{aligned}
\end{equation}
where sets $\mathcal{L}_{c,e}$ and $\mathcal{L}_{s,e}$ denote the detection error of communication and sensing tasks (including both detection and collision error), set $\mathcal{L}_{s,d}$ denotes the AOA of successfully detected SUs and $\hat{\theta}_k$ is the corresponding AOA estimate. The energy constraint $E/N_0$ is defined as \textit{energy-per-user}, i.e.,
\begin{equation}
	\frac{E}{N_0} = \frac{|\mathcal{A}_c|\bar{P}_cL+|\mathcal{A}_s|\bar{P}_sL}{\sigma^2\left(|\mathcal{A}_c|+|\mathcal{A}_s|\right)},
\end{equation}
where $\bar{P}_c$ and $\bar{P}_s$ are the transmission energy per channel use of CUs and SUs respectively. The objective at the BS is to reach a set of prescribed communication and sensing targets, i.e., PUPE and MSEAOA, with $E/N_0$ as low as possible.

\section{Proposition: Achievable Results}\label{sec:pro}
For the proposed FAS-UNISAC model, with the power constraints ($\bar{P}_c$, $\bar{P}_s$) and system metrics in \eqref{eq:1}, we have
\begin{subequations}
\begin{align}
\mathrm{PUPE}&\leq \epsilon_{cons}+\epsilon_{coll}+\epsilon_{md},\label{eq:6-4},\\
\Delta_\mathrm{MSEAOA}&\leq\sum_{K_{s}=0}^{|\mathcal{A}_{s}|}\sum_{K_{c}=0}^{|\mathcal{A}_{c}|}P_{K_{s},K_{c}}\frac{16\sigma^4_z\gamma_{\mathrm{max}}}{\bar{\lambda}^2M^4}\label{eq:6-5}.
\end{align}
\end{subequations}
Relevant definitions are listed as follows:
\begin{itemize}
\item $P_{cons}$: The probability that at least one CU/SU surpasses the power constraint $\bar{P}_c$/$\bar{P}_s$;
\item $P_{coll}$: Collision-derived error.
\item $P_{md}$: Detection error defined by
\begin{equation*}
P_{md}=\frac{\mathbb{E}\left \{|\mathcal{L}_{c,md}|+|\mathcal{L}_{s,md}|\right \}}{|\mathcal{A}_c|+|\mathcal{A}_s|},
\end{equation*}
where $K_c=|\mathcal{L}_{c,md}|$, $K_s=|\mathcal{L}_{s,md}|$ denote the number of errors for communication and sensing detection.
\item $P_{K_s,K_c}$: Joint probability of $K_c$ and $K_s$ numbers of communication and sensing errors, i.e.,
\begin{equation*}
P_{K_s,K_c}=\mathbb{P}(|\mathcal{L}_{c,md}|=K_c,|\mathcal{L}_{s,md}|=K_s).
\end{equation*}
\end{itemize}
The above quantities are given by
\begin{subequations}
\begin{align}
\epsilon_{cons}&=1-F_{\chi^2}\left(\frac{2L\bar{P}_s}{P_s^{\prime}},2L\right)^{|\mathcal{A}_s|}F_{\chi^2}\left(\frac{2L\bar{P}_c}{P_c^{\prime}},2L\right)^{|\mathcal{A}_c|}\label{eq:6-6},\\
\epsilon_{coll}&\leq\frac{\sum_{i=2}^\infty\frac{i\binom{|\mathcal{A}_s|}{i}}{2^{A_s(i-1)}}+\sum_{j=2}^\infty\frac{j\binom{|\mathcal{A}_c|}{j}}{2^{A_c(j-1)}}}{|\mathcal{A}_c|+|\mathcal{A}_s|}\label{eq:6-7},\\
\epsilon_{md}&=\sum_{K_s=0}^{|\mathcal{A}_s|}\sum_{K_c=0}^{|\mathcal{A}_c|}\frac{K_c+K_s}{|\mathcal{A}_c|+|\mathcal{A}_s|}P_{K_s,K_c}\label{eq:6-8},\\
P_{K_{s},K_{c}}&\leq e^{L_{s}+L_{c}-LM\log\left(1+0.25\sigma_{t}^{2}/\sigma^{2}\right)}\label{eq:6-9},\\
L_c&=\sum_{i=0}^{K_c-1}\log\left(\frac{|\mathcal{A}_c|-i}{K_c-i}\right)+\log\left(\frac{2^{{A_c}}-i}{K_c-i}\right)\label{eq:6-10},\\
L_s&=\sum_{i=0}^{K_s-1}\log\left(\frac{|\mathcal{A}_s|-i}{K_s-i}\right)+\log\left(\frac{2^{A_s}-i}{K_s-i}\right)\label{eq:6-11},\\
\sigma_t^2&=\frac{1}{M}\mathbb{E}\left\{\|\boldsymbol{g}_j\|_2^2\right\}K_cP_c^{\prime}+K_sP_s^{\prime}\label{eq:6-12},\\
\sigma_z^2&=\frac{\sigma^2+\sigma_t^2}{\|\boldsymbol{a}_j\|_2^2}\\
\bar{\lambda}^2&=\left(\frac{2\pi\mathbb{E}\{|N_m-N_n|\}W}{N_f-1}\right)^2 \label{eq:6-13}\\
\gamma_{\mathrm{max}}&=\|\boldsymbol{A}\|_2^2.\label{eq:6-14}
\end{align}
\end{subequations}
In \eqref{eq:6-6}, $P_c'$ and $P_s'$ denote the transmitted power of CUs and SUs abiding by $P_c'\le \bar{P}_c$, $P_s'\le \bar{P}_s$. In \eqref{eq:6-12}, $\frac{1}{M}\mathbb{E}\left\{\|\boldsymbol{g}_j\|_2^2\right\}, j\in [1:|\mathcal{A}_c|]$ denotes the averaged channel gains by the spatial diversity inherent within the fluid antenna. The corresponding value will be determined by Monte Carlo simulations. In \eqref{eq:6-13}, $\mathbb{E}\{|N_m-N_n|\}$ is the averaged difference between any two selected fluid antenna array elements, which can be found in Table \ref{tab:MRA pattern}. Finally, in \eqref{eq:6-14}, $\boldsymbol{A}$ denotes the sensing codebook in \eqref{Lasso} whose largest eigenvalue $\gamma_{\mathrm{max}}$ is determined by Monte Carlo simulations due to its random nature.

\section{Communication: Optimization and Analysis}\label{sec:com}
The error in (\ref{eq:6-4}) is derived from $P_{cons}$, $P_{colli}$ and $P_{md}$ as defined. We explain each type-of-error below. 

\begin{enumerate}
\item $\epsilon_{colli}$ and $\epsilon_{cons}$:
For any $x$ active users with a $N_c$-size common codebook, the averaged number of collisions from $x_c$ users on one codeword is given by
\begin{equation}
E_{(x_c,x,N_c)}=\frac{\binom{x}{x_c}\binom{N_c}{1} }{N_c^{k_c}}=\frac{\binom{x}{x_c} }{N_c^{k_c-1}},
\end{equation} 
producing the probability of collision error of
\begin{equation}
P(x_c,x,N_c) = \frac{E_{(x_c,x,N_c)}}{x}.
\end{equation}
Accumulating the error in all $x_c\ge 2$ users, we have
\begin{equation}
P_{e-colli}(x,N_c)= \textstyle \sum_{k_c=2}^{x}k_c \cdot P(x_c,x,N_c).
\end{equation}
As such, the collision error upper bound is found as
\begin{equation}
\epsilon_{coll}\leq\frac{\sum_{i=2}^\infty\frac{i\binom{|\mathcal{A}_s|}{i}}{2^{A_s(i-1)}}+\sum_{j=2}^\infty\frac{j\binom{|\mathcal{A}_c|}{j}}{2^{A_c(j-1)}}}{|\mathcal{A}_c|+|\mathcal{A}_s|}.
\end{equation}
Notably, recent work has revealed that certain collision case may not necessarily cause errors \cite{URA_TVT}. Thus, it is reasonable to treat \eqref{eq:6-7} as a collision error upper bound.
 
For error $P_{cons}$, a random codebook $\boldsymbol{A}$ with elements in the first $2^{A_c}$ rows and the last $2^{A_s}$ rows is, respectively, drawn from $\mathcal{CN}(0,P_c')$ and $\mathcal{CN}(0,P_s')$. By definition, $P_{cons}$ can be obtained by
\begin{equation}
\epsilon_{cons}=1-\prod_{j\in\{\mathcal{A}_c,\mathcal{A}_s\}}\mathbb{P}\left(\|\boldsymbol{a}_i\|^2/L<\bar{P}_l\right),
\end{equation}
where $\frac{2}{P_l'}\|\boldsymbol{a}_i\|_2^2\sim\chi_{2L}^2, l\in \{c,s\}$. By the assumption of chi-square distribution, the result in \eqref{eq:6-6} can be deduced. 
 
\item $\epsilon_{md}$: For detection error $\epsilon_{md}$, the existing model is briefly introduced and the spatial diversity brought by fluid antenna is explained. In accordance with \cite{ISAC-URA}, the codeword detection can be written as
\begin{equation}\label{detection_model}
\hat{\boldsymbol{A}}_d=\arg\min_{\boldsymbol{A}_d}\|\boldsymbol{Y}f_p\left(\boldsymbol{A}_d\right)\|^2,
\end{equation}
where $\boldsymbol{A}_d\in \mathbb{C}^{|\mathcal{A}_c|+|\mathcal{A}_s|\times L}$ appends all active codewords to be detected with $\binom{2^{A_s}}{\left|\mathcal{A}_s\right|}\binom{2^{A_c}}{\left|\mathcal{A}_c\right|}$ types of combinations and the detection function is defined as $f_p\left(\boldsymbol{A}_d\right)=\boldsymbol{I}_L-\boldsymbol{A}_d^H(\boldsymbol{A}_d\boldsymbol{A}_d^H)^{-1}\boldsymbol{A}_d$. We rewrite the received signal \eqref{eq:6-1} into $\boldsymbol{Y}=\boldsymbol{G}_a\boldsymbol{A}_a+\boldsymbol{Z}$ and define the detection matrix with error as $\boldsymbol{A}_e=\begin{bmatrix}\boldsymbol{A}_{\mathrm{correct}}\\\boldsymbol{A}_{e}\end{bmatrix}\in \mathbb{C}^{(K_c+K_s)\times L}$ with $K_c$ communication errors and $K_s$ sensing errors. That is, if $\boldsymbol{A}_e$ is declared as the solution to \eqref{detection_model}, the missed detection error $P_{md}$ will be obtained. By accounting all erroneous combinations, one can get
\begin{equation}
\begin{aligned}
P_{K_{s},K_{c}}&=\mathbb{P}\left(\bigcup_{\boldsymbol{A}_{e}\in\mathcal{A}_{A}}\bigcap_{\boldsymbol{A}_{e}^{\prime}\in\Omega}\{\zeta_{\boldsymbol{A}_{e},\boldsymbol{A}_{e}^{\prime}}\}\right)\\
&\leq|\mathcal{A}_{A}|\mathbb{P}\left(\zeta_{\boldsymbol{A}_{e},\boldsymbol{A}_{a}}\right),
\end{aligned}
\end{equation}
where $\zeta_{\boldsymbol{A}_1,\boldsymbol{A}_2}=\{\|\boldsymbol{Y}f_p(\boldsymbol{A}_1)\|^2\leq\|\boldsymbol{Y}f_p(\boldsymbol{A}_2)\|^2\}$, $\Omega$ is the universal set and $\mathcal{A}_{A}$ includes all possible choices for $\boldsymbol{A}_e$ so that 
\begin{equation}\label{Error_Space}
|\mathcal{A}_A|=\begin{pmatrix}2^{A_s}\\K_s\end{pmatrix}\begin{pmatrix}|\mathcal{A}_s|\\K_s\end{pmatrix}\begin{pmatrix}2^{A_c}\\K_c\end{pmatrix}\begin{pmatrix}|\mathcal{A}_c|\\K_c\end{pmatrix},
 \end{equation}
 where the terms in \eqref{eq:6-10} and \eqref{eq:6-11} originate from $L_{s}+L_{c}=\log\left(\binom{2^{B_{s}}}{K_{s}}\binom{|\mathcal{A}_{s}|}{K_{s}}\binom{2^{B_{c}}}{K_{c}}\binom{|\mathcal{A}_{c}|}{K_{c}}\right)$. 

Following the elegant derivations in \cite{ISAC-URA}, one can express $P_{K_s,K_c}$ into the function of averaged variance of channel elements, given by
\begin{equation}
\label{error_prob}
\begin{aligned}
P_{K_{s},K_{c}}&\leq e^{L_{s}+L_{c}-LM\log\left(1+0.25\sigma_{t}^{2}/\sigma^{2}\right)},\\
\sigma_t^2&=\frac{1}{M}\mathbb{E}\left\{\|\boldsymbol{g}_j\|_2^2\right\}K_cP_c^{\prime}+K_sP_s^{\prime},
\end{aligned}
\end{equation}
which reveals that $P_{K_s,K_c}$ is inversely proportional to the averaged variance of channel response, i.e., $\frac{1}{M}\mathbb{E}\left\{\|\boldsymbol{g}_j\|_2^2\right\}, j\in [1:|\mathcal{A}_c|]$. The derivation of \eqref{error_prob} is given in Appendix \ref{app:1}.

In terms of the array response within a fluid antenna, the fluctuation of array responses takes place when the LOS/NLOS channel components overlap. By assuming \textit{optimal ports selection}, the $M$ array elements with the highest channel responses can be sorted out of the $N_f$ available ports by
\begin{equation}\label{ports_selection}
\boldsymbol{g}_j=\mathop{\arg\max}_{\text{$M$ Largest Ones}}\{|g_{j,1}|,|g_{j,2}|,\ldots,|g_{j,N_f}|\}\in \mathbb{C}^{M\times 1}.
\end{equation}
Different from the channel model with fluid antenna, the element-wise modulus of the LOS-only channel model does not fluctuate by location and remains identical to unit modulus. In other words, the spatial diversity from the fluid antenna can produce an array response gain, i.e., the activated ports will presumably have $\|\boldsymbol{g}_j\|_2^2\ge \|\boldsymbol{g}_{\text{LOS-Only}}\|_2^2$ if there are array response gains. 
\end{enumerate} 

\section{Sensing: Optimization and Analysis}\label{sec:Sen}
In this section, we establish the sensing upper bound for AOA estimation in the FAS-UNISAC model. We first verify that the existing pessimistic estimation model is applicable under the irregular sensing array model of FAS-UNISAC. The assumption of high level interference under the pessimistic model naturally leads to an MSEAOA upper bound. Based on the pessimistic assumption, the model optimization by the spatial diversity of fluid antenna is explained along with an analytical solution to the optimized sensing model.

\subsection{Scalability of Pessimistic Sensing Model}
To upper bound the sensing estimation error for the proposed FAS-UNISAC system, we can verify the scalability of the pessimistic ULA estimation model in \cite[(20)]{ISAC-URA}, where the AOA estimation model can be expressed as
\begin{equation}\label{AOA_model}
\begin{aligned}
\hat{\boldsymbol{g}}_{\theta_i}&=\boldsymbol{g}_{\theta_i}+\boldsymbol{z}_i,\\
\boldsymbol{z}_i&\sim \mathcal{CN}\left(0,\sigma_z^2=\frac{\sigma^2+\sigma_t^2}{\|\boldsymbol{a}_j\|_2^2}\right), j\in \mathcal{A}_s,\\
\end{aligned}
\end{equation}
where $\hat{\boldsymbol{g}}_{\theta_i}^T\hat{\boldsymbol{g}}_{\theta_i}^R=\boldsymbol{g}_{\theta_i}^T\boldsymbol{g}_{\theta_i}^R+2\boldsymbol{g}_{\theta_i}^T\boldsymbol{z}_i^R+\boldsymbol{z}_i^T\boldsymbol{z}_i^R$.

For ULA with half-wavelength spacing and $M$ uniformly divided elements, we have
\begin{equation}\label{T-R1}
\begin{aligned}
&\begin{bmatrix}
	1,e^{-j\pi \cos\theta_i},
	\dots,e^{-j\pi(M-1) \cos\theta_i}\end{bmatrix} \cdot
	\begin{bmatrix}
		e^{-j\pi(M-1)\cos\theta_i}\\
		e^{-j\pi (M-2)\cos\theta_i}\\
		\vdots\\
		1
	\end{bmatrix}\\
	&=Me^{-j\pi(M-1)\theta_i}.
	\end{aligned}
\end{equation}
By contrast, for the FAS model, the integers in the exponentiation terms can be randomized by different array elements selection strategies, i.e.,
\begin{equation}\label{T-R2}
\begin{aligned}
&\begin{bmatrix}
			e^{-j\frac{2\pi N_1 W}{N_f-1}\cos\theta_i},
			\dots,
			e^{-j\frac{2\pi N_M W}{N_f-1}\cos\theta_i}
			\end{bmatrix} \cdot\begin{bmatrix}
			e^{-j\frac{2\pi N_M W}{N_f-1}\cos\theta_i}\\
			\vdots\\
		e^{-j\frac{2\pi N_1 W}{N_f-1}\cos\theta_i}
		\end{bmatrix}\\
&=\sum_{i=1}^{i=M}e^{-j\frac{2\pi(N_i+N_{M+1-i})W}{N_f-1}\cos\theta_i} = M e^{-j2W\pi\cos\theta_i},
\end{aligned}
\end{equation}
where by the law of large number, $N_i\in [0:N_f-1]$ can be treated as a random variable with an expected mean of $\frac{1}{N_f}\left(0+1+2+\ldots+(N_f-1)\right)=\frac{N_f-1}{2}$, i.e., $\mathbb{E}\{N_i+N_{M+1-i}\}=N_f-1$.

As a result, the receiver estimates the AOA $\theta_i$ based on different results of $\boldsymbol{g}_{\theta_i}^T\boldsymbol{g}_{\theta_i}^R$ in \eqref{T-R1} and \eqref{T-R2}, yielding
\begin{equation}\label{AOA_error}
\begin{aligned}
&\mathrm{ULA \ by\ \eqref{T-R1}:} \quad \hat{\theta}_i=\mathrm{Re}\left(\frac{1}{-j\pi(M-1)}\log\left(\frac{1}{M}\hat{\boldsymbol{g}}_{\theta_i}^T\hat{\boldsymbol{g}}_{\theta_i}^R\right)\right)\\
&=\theta_i+\\&\underbrace{\mathrm{Re}\left(\frac{1}{-j\pi(M-1)}\operatorname{log}(1+\frac{e^{j\pi(M-1)\cos\theta_{i}}}{M}(2\boldsymbol{g}_{\theta_i}^T\boldsymbol{z}_i^R+\boldsymbol{z}_i^T\boldsymbol{z}_i^R))\right)}_{\text{Estimation Deviation}}\\
&\mathrm{FAS \ by \ \eqref{T-R2}:} \quad
\hat{\theta}_i=\mathrm{Re}\left(\frac{1}{-j2W\pi}\log\left(\frac{1}{M}\hat{\boldsymbol{g}}_{\theta_i}^T\hat{\boldsymbol{g}}_{\theta_i}^R\right)\right)\\
&=\theta_i+\underbrace{\mathrm{Re}\left(\frac{1}{-j2W\pi}\operatorname{log}(1+\frac{e^{j2W\pi\cos\theta_{i}}}{M}(2\boldsymbol{g}_{\theta_i}^T\boldsymbol{z}_i^R+\boldsymbol{z}_i^T\boldsymbol{z}_i^R))\right)}_{\text{Estimation Deviation}}.
\end{aligned}
\end{equation}

FAS and ULA follow the identical constraint of antenna size, i.e., $W=\frac{(M-1)}{2}$. One can observe the same expression of estimation deviation among ULA and FAS when the LOS-only channel is considered with a random shifting factor, i.e., the terms $e^{j\pi (M-1)\cos\theta_i}$ and $e^{j2W\pi \cos\theta_i}$ in \eqref{AOA_error} do not affect distribution. The pessimistic estimation model assumption upper-bouds the estimation error with high level of interference, based on which the optimization on AOA estimation is made with the spatial diversities in fluid antenna. 

\subsection{Sensing Optimization Model}
In this subsection, we discuss the optimization based on the pessimistic model via the spatial diversity of fluid antenna. More importantly, an estimation error upper bound is derived. By the model \eqref{AOA_model} and assuming independence between noise and the sensing signal, the covariance matrix of noisy observation $\hat{\boldsymbol{\Phi}}\in \mathbb{C}^{M\times M}$ can be written by
\begin{equation}
\begin{aligned}
\hat{\boldsymbol{\Phi}}&=\mathbb{E}\left[\hat{\boldsymbol{g}}_{\theta_i}\hat{\boldsymbol{g}}^{\mathrm{H}}_{\theta_i}\right]=\boldsymbol{\Phi}+\sigma_z^2\boldsymbol{I}_M\\
&=\boldsymbol{g}_{\theta_i}\boldsymbol{g}^{\mathrm{H}}_{\theta_i}+\sigma_{z}^2\boldsymbol{I}_M,
\end{aligned}
\end{equation}
where $\boldsymbol{\Phi}\in \mathbb{C}^{M\times M}$ is the covariance matrix of signal component given by
\begin{equation}
	\begin{aligned}
	\boldsymbol{\Phi}=&\begin{bmatrix}
		e^{-j\frac{2\pi N_1 W}{N_f-1}\cos\theta_i},
		\dots,
		e^{-j\frac{2\pi N_M W}{N_f-1}\cos\theta_i}
	\end{bmatrix} \cdot\begin{bmatrix}
		e^{j\frac{2\pi N_1 W}{N_f-1}\cos\theta_i}\\
		\vdots\\
		e^{j\frac{2\pi N_M W}{N_f-1}\cos\theta_i}
	\end{bmatrix}\\
	&\rightarrow \left[\boldsymbol{\Phi}\right]_{m,n}=e^{-j\underbrace{(N_m-N_n)}_{\text{Index Difference}}\frac{2\pi W\cos\theta_i}{N_f-1}}, m,n\in [1:M].
\end{aligned}
\end{equation}
Thus, the element at the $m$-th row and the $n$-th column of $\hat{\boldsymbol{\Phi}}$ is found as
\begin{equation}
\left[\hat{\boldsymbol{\Phi}}\right]_{m,n} =e^{-j\underbrace{(N_m-N_n)}_{\text{Index Difference}}\frac{2\pi W\cos\theta_i}{N_f-1}}+\sigma_z^2\delta_{m,n},
\end{equation}
where $\delta_{m,n}$ is the Kronecker delta function with $\delta_{m,n}=1$ if $m=n$, otherwise $\delta_{m,n}=0$. Notably, the factor $(N_m-N_n)$ in the exponential term is the difference between the randomized indices of array elements within the fluid antenna, i.e., $N_m, N_n\in [0:N_f-1]$. Based on this observation, we define a difference co-array (DCA) \cite{DCA} as
\begin{equation}
\mathcal{N}=\left \{N_m-N_n,\forall N_m,N_n \in [0:N_f-1]\right \}.
\end{equation}
By reasonably selecting different  $M$ elements to be activated, one can obtain a subset $\mathcal{N}_{\mathrm{sub}} \subseteq \mathcal{N}$ with subset cardinality (number of unique elements) much larger than the number of activated array elements. Thus, more DOF can be generated utilizing the second-order statistic feature from the covariance matrix. By matrix vectorization on $\hat{\boldsymbol{\Phi}}$, the original sensing signal model in \eqref{AOA_model} can be converted into a virtual sensing signal model $\boldsymbol{v}\in \mathbb{C}^{M^2\times 1}$ with DCA elements in $\mathcal{N}_{\mathrm{sub}}$:
\begin{equation}\label{DCA-AOA}
\begin{aligned}
\boldsymbol{v}&=\mathrm{vec}(\hat{\boldsymbol{\Phi}})=\mathrm{vec}(\boldsymbol{g}_{\theta_i}\boldsymbol{g}^{\mathrm{H}}_{\theta_i}+\sigma_{z}^2\boldsymbol{I})\\
&=\underbrace{\boldsymbol{g}_{\theta_i}^*\otimes \boldsymbol{g}_{\theta_i}}_{\text{virtual DCA array}}+\sigma_{z}^2\boldsymbol{I}_{M^2},
\end{aligned}
\end{equation}
in which $\otimes$ represents the Kronecker product and $\boldsymbol{I}_{M^2}=\left[\boldsymbol{e}_1^{\mathrm{T}},\boldsymbol{e}_2^{\mathrm{T}},\ldots, \boldsymbol{e}_M^{\mathrm{T}}\right]$ and column vector $\boldsymbol{e}_i,i\in [1:M]$ refers to the indicator vector with only one non-zero element equal to $1$ at the $i$-th index and others all zeros. Hence, the virtual AOA estimation model in \eqref{DCA-AOA} has a \textit{wider receiving aperture} than the original model in \eqref{AOA_model}, i.e., $M^2> M$.

Subsequently, how to appropriately design the DCA subset $\mathcal{N}_{\mathrm{sub}}$ is explained. Let $x=(N_m-N_n)$ denote the difference between two random activated array indices and let the weight function $w(x)$ denote the number of a pair of array indices with difference equal to $x$. For any $x\in \mathcal{N}_{\mathrm{sub}}$, $w(x)$ should incorporate the following features \cite{DCA}:
\begin{subequations}
	\label{f}
	\begin{align}
		&w(0)=M,
		\label{f_1},\\
		&\forall x\in \mathcal{N}_{\mathrm{sub}}\setminus \{0\}, 1\le w(x) \le M-1
		\label{f_2},\\
		&\forall x\in \mathcal{N}_{\mathrm{sub}}, w(x)=w(-x),
		\label{f_3}\\
		&\sum_{x\in \mathcal{N}_{\mathrm{sub}},x\neq 0}w(x)=M(M-1)
		\label{f_4}.
	\end{align}
\end{subequations}

By using \eqref{f_4}, one can observe a maximum DOF of $\mathcal{N}_{\mathrm{sub}}$ in the order of $M(M-1)+1$. When $w(x)>1,\forall x\in \mathcal{N}_{\mathrm{sub}}\setminus \{0\}$, the DOF decreases. In other words, the selected $\mathcal{N}_u$ directly influences the DOF of the observed signal \eqref{DCA-AOA}. 

\subsection{Port Selection for FAS}
By the principle of minimum redundancy linear arrays (MRA) \cite{MRA1,MRA2}, one can maximize the DOF of $\mathcal{N}_{\mathrm{sub}}$ by searching the DCA with the least redundancy, i.e., finding the combinations on port activation so that $w(x)\rightarrow 1,\forall x\in \mathcal{N}_{\mathrm{sub}}\setminus \{0\}$. Though MRA does not have explicit algebraic solutions, one can obtain favourable ports via an exhaustive search to satisfy the constraints \eqref{f}. Let $\mathcal{P}$ denote the port selection index, i.e., $N_m,N_n\in \mathcal{L}$ and some results of $\mathcal{P}$ are listed in Table \ref{tab:MRA pattern} where the weight function $w(x)$ generated from $N_n,N_m\in \mathcal{P}$ satisfy all the four constraints in \eqref{f}.

\begin{table}[htp]
\centering
\caption{MRA-Based Fluid Antenna Port Selection}\label{tab:MRA pattern}
\renewcommand{\arraystretch}{1.3}
\begin{tabular}{c|l|l}
		\hline
		$M$ & Index of Ports $\mathcal{P}$ To Be Activated                                                     & $\mathbb{E}\{N_m-N_n\}$ \\ \hline
		3   & [0,1,3] or [0,2,3]                                                                               & 1.3333                  \\ \hline
		5   & [0,1,4,7,9] or [0,1,2,6,9]                                                                       & 3.84 or 3.68            \\ \hline
		7   & \begin{tabular}[c]{@{}l@{}}[0,1,2,6,10,14,17] \\ or [0,1,2,3,8,13,17]\end{tabular}               & 6.9388 or 6.6122        \\ \hline
		9   & \begin{tabular}[c]{@{}l@{}}[0,1,2,14,18,21,24,27,29]\\ or [0,1,3,10,16,22,24,27,29]\end{tabular} & 12.0988 or 12.2469      \\ \hline
		10  & [0,1,3,6,13,20,27,31,35,36]                                                                      & 15.44                   \\ \hline
		11  & [0,1,3,6,13,20,27,34,38,42,43]                                                                   & 18.314                  \\ \hline
	\end{tabular}
\end{table}

After virtual DCA array conversion, the coherent signal in \eqref{DCA-AOA} can not be solved directly by sub-space search methods such as multiple signal classification (MUSIC). \textit{Though one can adopt the sub-space search method after the smoothing method \cite{DCA} or matrix reconstruction method \cite{Toeplits}, there will be DOF loss after smoothing and matrix reconstruction.} 

\subsection{Sensing Estimation Upper Bound}
By contrast, compressive sensing (CS)-based method \cite{DCA-CS,DCA-CS1} can maintain full DOF in possessing the virtual DCA array with an appropriate sensing codebook, which is favourable for an upper-bound search. Let $\boldsymbol{A}=\left[\tilde{\boldsymbol{g}}_{\theta_1},\tilde{\boldsymbol{g}}_{\theta_2},\ldots,\tilde{\boldsymbol{g}}_{\theta_N}\right]\in \mathbb{C}^{M^2 \times N}$ denote the sensing codebook where $\tilde{\boldsymbol{g}}_{\theta_n}=\boldsymbol{g}^*_{\theta_n}\otimes \boldsymbol{g}_{\theta_n},n\in [1:N]$ and $N$ is the number of AOA samples which should be adequately large to satisfy the MSEAOA target. We rewrite \eqref{DCA-AOA} into a sparse linear regression model as
\begin{equation}\label{DCA-AOA-CS}
\boldsymbol{v}=\boldsymbol{A}\boldsymbol{\beta}+\boldsymbol{n}_z,
\end{equation}
where $\boldsymbol{\beta}$ is the unknown coefficient vector to be estimated and noise $\boldsymbol{n}_z$ denotes the vectorization version of $\sigma_{z}^2\boldsymbol{I}_{M^2}$, i.e., $\boldsymbol{n}_z=\left[\boldsymbol{e}_1^{\mathrm{T}},\boldsymbol{e}_2^{\mathrm{T}},\ldots,\boldsymbol{e}_M^{\mathrm{T}}\right]^{\mathrm{T}}$ with only $M$ nonzero elements. The solution $\hat{\boldsymbol{\beta}}$ of \eqref{DCA-AOA-CS} should satisfy the following constraints:
\begin{equation}\label{NP-Hard}
\hat{\boldsymbol{\beta}}=\arg\min_{\boldsymbol{\beta}}\lVert\boldsymbol{\beta}\rVert_0\quad s.t.\quad\lVert\boldsymbol{v}-\boldsymbol{A\beta}\rVert_2^2<\varepsilon_0,
\end{equation}
where $\varepsilon_0$ is the prescribed deviation tolerance. While \eqref{NP-Hard} is NP-hard, a robust solution can be derived by converting $l_0$-norm into $l_1$-norm that is treated as a classic Lasso problem:
\begin{equation}\label{Lasso}
\hat{\boldsymbol{\beta}}=\underset{\|\beta\|_1\leq \varepsilon_1}{\operatorname*{\operatorname*{min}}}\|\boldsymbol{v}-\boldsymbol{A}\boldsymbol{\beta}\|_2^2,
\end{equation}
where in accordance with \cite[(11.14a), Chapter 11]{Lasso_Upper}, the estimation error upper bound of the 1-sparse vector $\hat{\boldsymbol{\beta}}$ can be expressed as
\begin{equation}\label{Lasso-Error-upper}
\|\boldsymbol{\beta}-\hat{\boldsymbol{\beta}}\|_2\le \frac{4}{M}\left\|\frac{\boldsymbol{A}^{\mathrm{H}}\boldsymbol{n}_z}{M}\right\|_{\infty},
\end{equation}
which is derived in Appendix \ref{app:2}.

The $l_\infty$-norm at the right side can be obtained by
\begin{equation}\label{inf}
\left\|\frac{\boldsymbol{A}^{\mathrm{H}}\boldsymbol{n}_z}{M}\right\|_{\infty}=\max_{i\in[1:N]}\left|\frac{1}{M}\tilde{\boldsymbol{g}}_{\theta_i}^{\mathrm{H}}\boldsymbol{n}_z\right|=\sigma^2_z.
\end{equation}
Noticeably, the vectorized noise from its covariance matrix has only two elements, i.e., $\{0,\sigma^2\}$ with only $M$ nonzero elements and the modulus of elements in any sensing codeword equals to $1$. Thus, we have $\max_{i\in[1:N]}|\frac{1}{M}\tilde{\boldsymbol{g}}_{\theta_i}^{\mathrm{H}}\boldsymbol{n}_z|=\sigma^2_z$.

Subsequently, we bridge the MSEAOA with Lasso error bound derived in \eqref{Lasso-Error-upper}. The averaged estimation deviation on the virtual DCA element can be written by
\begin{equation}\label{DCA-Element-Estimation}
\begin{aligned}
&\mathbb{E}\left\{
\left |e^{-j\frac{2\pi(N_m-N_n)W}{N_f-1}\cos\theta_{i} }-e^{-j\frac{2\pi(N_m-N_n)W}{N_f-1}\cos\hat{\theta}_{i} }\right|^2
\right\}\\
&=\frac{1}{M^2}\|\boldsymbol{A\beta}-\boldsymbol{A}\hat{\boldsymbol{\beta}}\|_2^2\le \frac{1}{M^2}\|\boldsymbol{A}\|_2^2\|\boldsymbol{\beta}-\hat{\boldsymbol{\beta}}\|^2_2.
\end{aligned}
\end{equation}
For ease of description, we alternate the factor term at the exponent by $\lambda=\frac{2\pi(N_m-N_n)W}{N_f-1}$. Then, the left-side term in \eqref{DCA-Element-Estimation} can be approximated by
\begin{equation}\label{Taylor Expansion}
\begin{aligned}
		&\mathbb{E}\left\{
		\left |e^{-j\lambda\cos\theta_{i} }-e^{-j\lambda\cos\hat{\theta}_{i} }\right|^2
		\right\}\\
		&=\mathbb{E}\left\{
		\left |e^{-j\lambda\cos\hat{\theta}_{i}}\left(e^{-j\lambda\left(\cos\theta_{i}-\cos\hat{\theta}_{i}\right)}-1 \right)\right|^2
		\right\}\\
		&=\mathbb{E}\left\{
		\left |\left(e^{-j\lambda\left(\cos\theta_{i}-\cos\hat{\theta}_{i}\right)}-1 \right)\right|^2
		\right\}\\
		&\xrightarrow[]{\text{Taylor Expansion}}
		\mathbb{E}\left\{
		\left |-j\lambda\left(\cos\theta_{i}-\cos\hat{\theta}_{i}\right)\right|^2
		\right\}\\
		&=\bar{\lambda}^2\mathbb{E}\left\{
		\left |\cos\theta_{i}-\cos\hat{\theta}_{i}\right|^2,
		\right\},
\end{aligned}
\end{equation}
where $\bar{\lambda}^2=\left(\frac{2\pi\mathbb{E}\{|N_m-N_n|\}W}{N_f-1}\right)^2$ and $\mathbb{E}\{|N_m-N_n|\}$ is listed in Table \ref{tab:MRA pattern}. Substituting \eqref{Lasso-Error-upper}, \eqref{inf} and the approximation of \eqref{Taylor Expansion} into \eqref{DCA-Element-Estimation}, we obtain
\begin{equation}\label{MSEAOA_Upper_Bound}
\begin{aligned}
\mathbb{E}\left\{\left |\cos\theta_{i}-\cos\hat{\theta}_{i}\right|^2\right\}&\le \frac{1}{\bar{\lambda}^2M^2}\|\boldsymbol{A}\|_2^2\|\boldsymbol{\beta}-\hat{\boldsymbol{\beta}}\|^2_2\\
&\le\frac{16\sigma^4_z\gamma_{\mathrm{max}}}{\bar{\lambda}^2M^4},
\end{aligned}
\end{equation}
where $\|\boldsymbol{A}\|_2^2=\gamma_{\mathrm{max}}$ and $\gamma_{\mathrm{max}}$ is the largest eigenvalue of $\boldsymbol{A}^{\mathrm{H}}\boldsymbol{A}$. Thus, the upper bound of MSEAOA is obtained.

\section{Performance Lower Bound}\label{sec:lower}
In this section, we obtain the lower bound for the FAS-UNISAC system. The lower bound of the performance is termed as the \textit{optimistic bound} due to ideal assumptions to obtain the potential minimum errors including MSEAOA and PUPE. Overall, the optimistic bound only considers collision-derived error $\epsilon_{colli}$ and use the single-user Cramer-Rao lower bound (CRLB) \cite{CRLB} as the sensing estimation lower bound. Under certain level of energy-per-user $E/N_0$ representing the performance lower bound, we have the system metrics:
\begin{subequations}\label{ob1}
\begin{align}
\text{PUPE}&\leq \sum_{l\in\{c,s\}}\frac{\binom{|\mathcal{A}_l|}{2}}{2^{B_l}}\left(\frac{2^{B_l}-1}{2^{B_l}}\right)^{|\mathcal{A}_l|-2}\label{ob_1},\\
\text{MSEAOA}&\leq \frac{0.5\sigma^2}{\pi^2L\bar{P}_s\sum_{i=1}^{M-1}i^2},\label{ob_2}
\end{align}
\end{subequations}
with the constraints of
\begin{subequations}\label{ob2}
\begin{align}
&\mathbb{E}\left\{\log_2\left(\det\left(\boldsymbol{I}_M+\frac{1}{\sigma^2}\boldsymbol{G}_{\text{FAS}}\boldsymbol{\Psi}\boldsymbol{G}_{\text{FAS}}^\mathrm{H}\right)\right)\right\}\rightarrow\frac{B_T}{L},\label{ob_3}\\
&B_T = A_c|\mathcal{A}_c|+A_s|\mathcal{A}_s|,\label{ob_4}
\end{align}
\end{subequations}
where $\boldsymbol{G}_{\text{FAS}}\in \mathbb{C}^{M\times |\mathcal{A}_c|+|\mathcal{A}_s|}$ is the channel matrix of all users and $\boldsymbol{\Psi}$ is a diagonal matrix with elements equal to $\bar{P}_c$ and $\bar{P}_s$ corresponding to the channel vector. Constraint \eqref{ob_3} is the averaged channel capacity/sum-rate at the multi-antenna system. According to Shannon's theorem, error-free transmission is possible when the rate stays below the averaged capacity. PUPE in \eqref{ob_1} incorporates only collision-derived error and MSEAOA in \eqref{ob_1} denotes the single-user CRLB. Notably, the number of bits transmitted is calculated by the capacity constraint in \eqref{ob_3} and \eqref{ob_4}.

\begin{figure}[htp]
\centering
\includegraphics[width=3.8in]{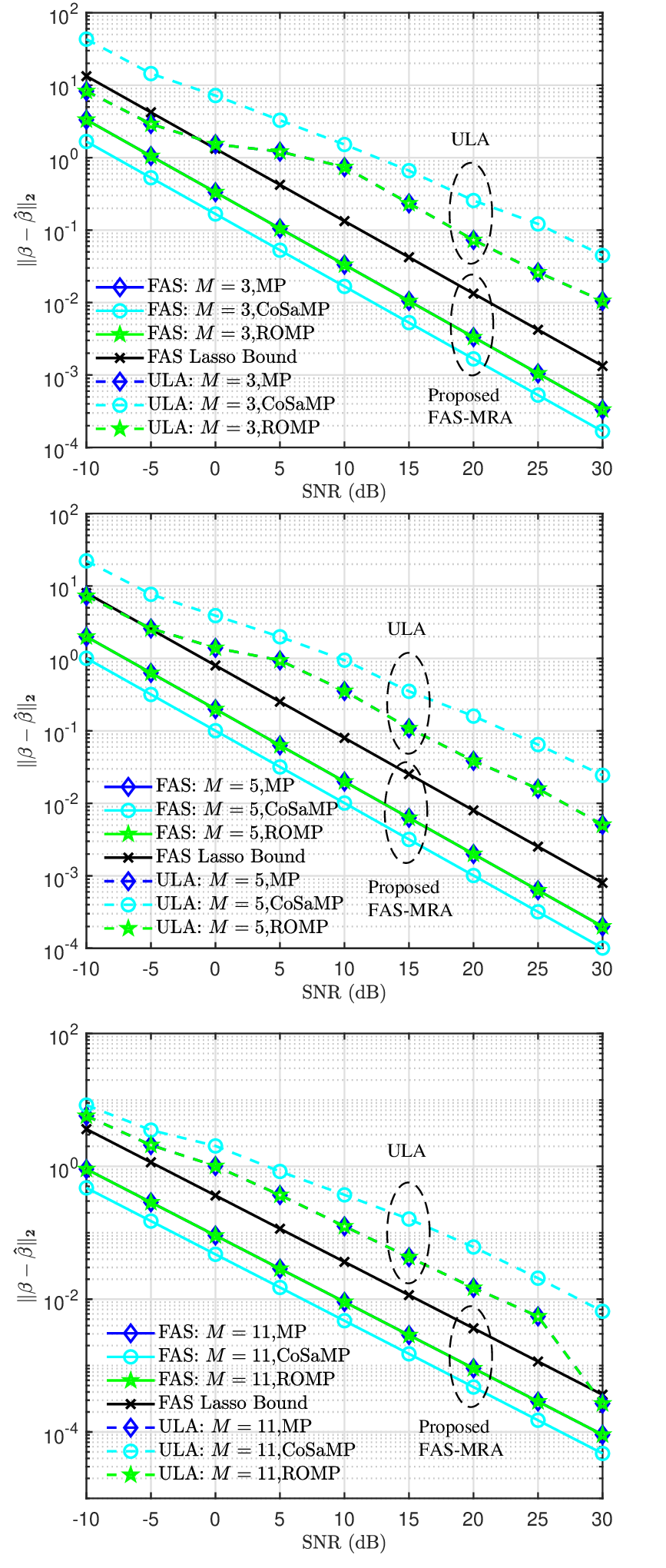}
\caption{The comparison between the proposed analytical CS estimation upper bound in \eqref{Lasso-Error-upper} and practical CS algorithms under channel model (FAS and ULA), different SNR (dB) and different number of receiving antennas $M\in\{3,5,11\}$.}\label{sim:CS}
\end{figure}

\section{Numerical Results}\label{sec:sim}
This section provides the numerical results of the proposed FAS-UNISAC with benchmark from the state-of-the-arts \cite{ISAC-URA}. System metrics under various parameter setups are simulated to demonstrate the superiority of fluid antenna for UNISAC. For systematic performance targets, PUPE is no more than $0.1$ and MSEAOA is under $5\times 10^{-4}$. In terms of fluid antenna channel models in \eqref{eq:6-2}, the Rice factor is fixed to $K=0.5$, single LOS path along with $L_s=3$ scatterer paths. The number of available ports equals to $|\mathcal{P}|$ and the ports activation pattern is listed in Table \ref{tab:MRA pattern}. For sensing codebook, the number of samples are fixed to $N=90$. The number of receiving antenna and channel uses are fixed to $M=10$ and $L=5000$, respectively, unless mentioned otherwise.

\subsection{Verification for $\|\boldsymbol{\beta}-\hat{\boldsymbol{\beta}}\|_2\le \frac{4}{M}\|\frac{\boldsymbol{A}^{\mathrm{H}}\boldsymbol{n}_z}{M}\|_{\infty}$}
The upper-bound estimation of sensing performance based on the proposed FAS-UNISAC model is essential to the final results. In Appendix \ref{app:2}, the proof on Lasso solutions in \eqref{Lasso-Error-upper} is given. Furthermore, numerical results by multiple prevalent CS algorithms are provided to demonstrate the correctness of $\|\boldsymbol{\beta}-\hat{\boldsymbol{\beta}}\|_2\le \frac{4}{M}\|\frac{\boldsymbol{A}^{\mathrm{H}}\boldsymbol{n}_z}{M}\|_{\infty}$. We first define the signal-to-noise ratio (SNR) based on model $\hat{\boldsymbol{g}}_{\theta_i}=\boldsymbol{g}_{\theta_i}+\boldsymbol{z}_i$ \eqref{AOA_model} by
\begin{equation}
\mathrm{SNR}=\frac{\mathbb{E}\left \{\|\boldsymbol{g}_{\theta_i}\|_2^2\right \}}{\mathbb{E}\left \{\|\boldsymbol{z}_i\|_2^2\right \}}=\frac{M}{M\sigma^2_z}=\frac{1}{\sigma_z^2}.
\end{equation}
Following the conversion from \eqref{AOA_model} into the virtual array model in \eqref{DCA-AOA-CS}, various CS algorithms are adopted to find solutions to the $\boldsymbol{\beta}$ and the averaged $\|\boldsymbol{\beta}-\hat{\boldsymbol{\beta}}\|_2$ are illustrated under different setups in comparison with the anticipated CS estimation upper bound in \eqref{Lasso-Error-upper} and \eqref{inf}.

In Fig.~\ref{sim:CS}, algorithms including the matching pursuit (MP) \cite{MP}, the compressive sampling matching pursuit (CoSaMP) \cite{CoSaMP}, and the regularized orthogonal matching pursuit (ROMP) \cite{ROMP} are conducted for comparison with the analytical upper bound $\|\boldsymbol{\beta}-\hat{\boldsymbol{\beta}}\|_2\le \frac{4}{M}\|\frac{\boldsymbol{A}^{\mathrm{H}}\boldsymbol{n}_z}{M}\|_{\infty}$. The ports activation patterns are selected from Table \ref{tab:MRA pattern} and the sensing matrix $\boldsymbol{A}$ is constructed correspondingly. While the estimation precision $\|\boldsymbol{\beta}-\hat{\boldsymbol{\beta}}\|_2$ is different among practical algorithms, they are all upper-bounded by the analytical results. Meanwhile, the CS-based AOA estimation is also simulated under the ULA channel model via codebook with identical sample size.

The results in Fig.~\ref{sim:CS} verify the prediction from the proof in Appendix \ref{app:2}. Note that the FAS channel simulation results of Lasso solutions by different practical algorithms are upper-bounded by the Lasso bound by $\|\boldsymbol{\beta}-\hat{\boldsymbol{\beta}}\|_2\le \frac{4}{M}\|\frac{\boldsymbol{A}^{\mathrm{H}}\boldsymbol{n}_z}{M}\|_{\infty}$ and thus the results validate the proposed FAS-UNISAC achievablility analysis deduced from the Lasso model. Interestingly, even the analytical upper bound results are lower than the ULA-codebook based estimation, which indicates the promising future of high-resolution FAS.

\begin{figure}[htp]
\centering
\includegraphics[width=3.5in]{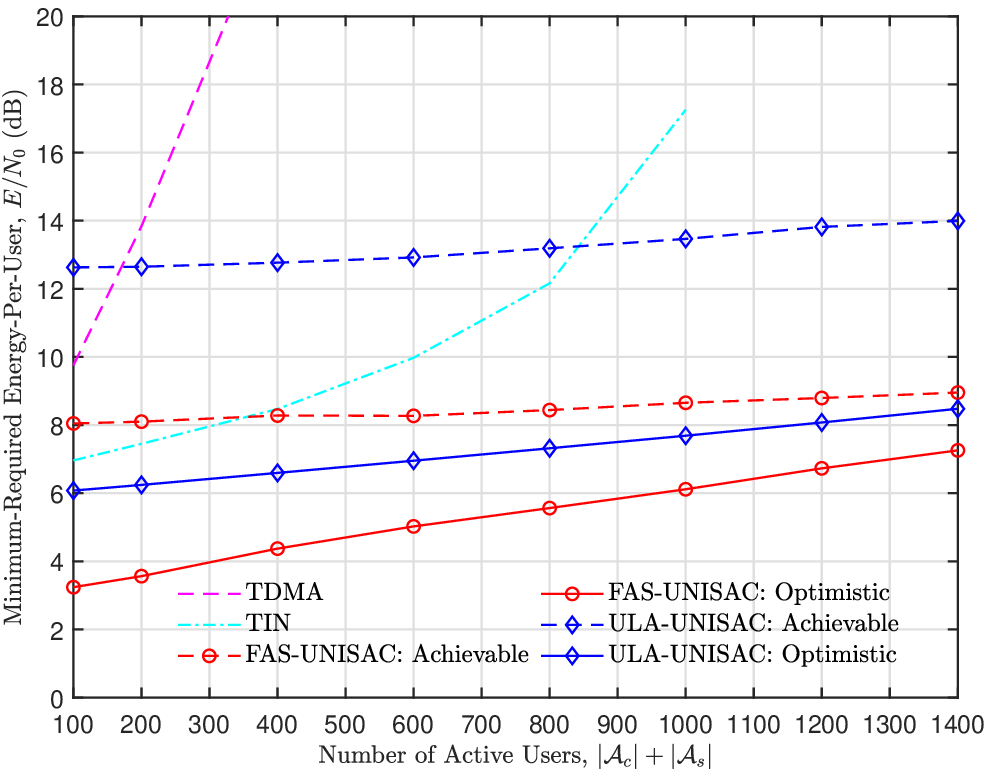}
\caption{Achievable results of FAS-UNISAC compared with benchmarks (UNISAC, TDMA, TIN) \cite{ISAC-URA} under $M=10$ receiving antenna and $L=5000$ channel uses: The performance target includes $0.1$ PUPE and $5\times 10^{-4}$ MSEAOA.}\label{sim:capacity}
\end{figure}

\subsection{Achievable Results}
Here, the achievable results and the performance lower bound are illustrated and compared with the existing state-of-the-art \cite{ISAC-URA}, i.e., the benchmarks are termed as LOS-UNISAC, TMDA and TIN. In Fig.~\ref{sim:capacity}, both achievable performance and performance lower bound (optimistic) are illustrated and compared with ULA-FAS. It can be observed that the proposed FAS-UNISAC has the lowest achievable results compared to both coordinated and uncoordinated multiple access schemes. At $1400$ active users, the achievable results of the proposed model are very close to the performance lower bound of LOS-UNISAC. Both FAS-UNISAC and UNISAC manifest great capacity in terms of the increasing number of users while conventional schemes (TDMA and TIN) will be overwhelmed. Overall, fluid antenna greatly improves the system performance via the spatial diversity within the array. Additionally, the gap between the achievable bound and the performance floor (optimistic) is tighter than the LOS-UNISAC, which is good in terms of performance evaluation. 

\begin{figure}[htp]
\centering
\includegraphics[width=3.5in]{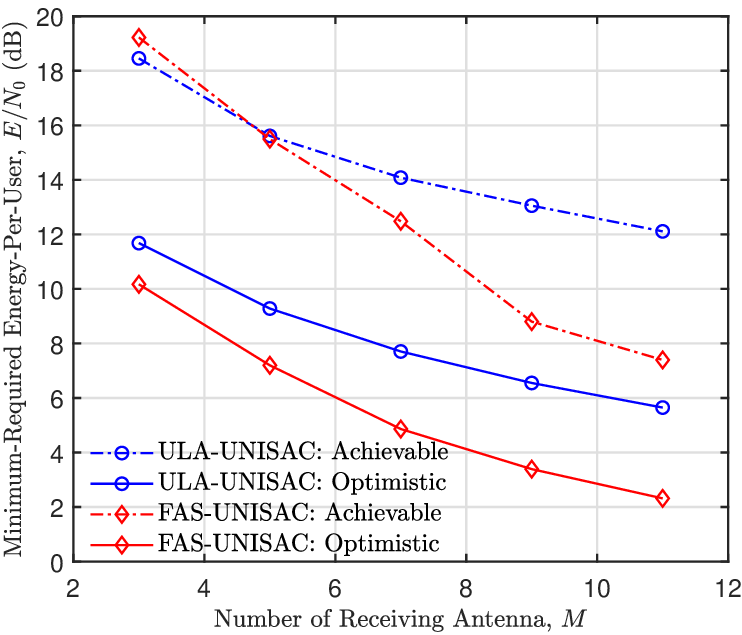}
\caption{Achievable results of FAS-UNISAC compared with benchmark \cite{ISAC-URA} under $|\mathcal{A}_c|+|\mathcal{A}_s|=100$ and different number of receiving antenna $M\in \{3,5,7,9,11\}$: The performance target includes $0.1$ PUPE and $5\times 10^{-4}$ MSEAOA.}\label{sim:capacity_M}
\end{figure}

Moreover, we also compare the minimum-required energy-per-user $E/N_0$ (dB) under a fixed number of users $|\mathcal{A}_c|+|\mathcal{A}_s|=100$ and different number of receiving antennas in Fig.~\ref{sim:capacity_M}. Explicitly, the required $E/N_0$ (dB) of the proposed FAS-UNISAC drops much faster than LOS-UNISAC with an increasing number of antennas. However, the proposed FAS-UNISAC model has lower required energy-per-user among all size of receiving antennas except for that when $M=3$, FAS-UNISAC has slightly higher required $E/N_0$, which can be derived from the sensing codebook. As revealed in \eqref{MSEAOA_Upper_Bound}, i.e., $\mathbb{E}\left\{\left |\cos\theta_{i}-\cos\hat{\theta}_{i}\right|^2\right\} \le\frac{16\sigma^4_z\gamma_{\mathrm{max}}}{\bar{\lambda}^2M^4}$, the MSEAOA is upper-bounded by $\gamma_{\mathrm{max}}$ which is the largest eigenvalue of the sensing codebook, i.e., $\gamma_{\mathrm{max}}=\|\boldsymbol{A}\|_2^2$. However, the issue is that $\gamma_{\mathrm{max}}$ increases exponentially with $\frac{N}{M^2}$, i.e., $\gamma_{\mathrm{max}}$ is extremely large when $M$ is relatively small, e.g., for the case in Fig.~\ref{sim:capacity_M}, an example is $\gamma_{\mathrm{max}}=1.5471\times 10^3$ with $M=3$ and $N=90$. High eigenvalue causes loss on performance. Though the loss is marginal in our case, how to design an optimal sensing codebook remains an open, and challenging problem.

\section{Conclusions}\label{sec:conc}
In this paper, the novel FAS-UNISAC system model has been proposed for mMTC applications. Our results have shown superiority of the proposed FAS-UNISAC model in comparison with the state-of-the art and the conventional schemes such as TDMA and TIN strategies. Through the spatial diversity of fluid antenna, the exploitation on the covariance information from sensing signals effectively expands the receiving aperture and thus greatly enhances the sensing resolution. Moreover, the Lasso sensing upper bound derived is also presented as a universal benchmark for relevant designs. 


\appendices
\section{Derivations on $P_{K_s,K_c}$} \label{app:1}
Here, we derive the relationship between the joint error probability $P_{K_s,K_c}$ and the averaged channel variance $\frac{1}{M}\mathbb{E}\left\{\|\boldsymbol{g}_j\| \right\}$ in accordance with the derivations in \cite{ISAC-URA}. By definition, the calculation $P_{K_s,K_c}$ can be done by
\begin{equation}
P_{K_{s},K_{c}}=\mathbb{P}\left(\bigcup_{\boldsymbol{A}_{e}\in\mathcal{A}_{A}}\bigcap_{\boldsymbol{A}_{e}^{\prime}\in\Omega}\{\zeta_{\boldsymbol{A}_{e},\boldsymbol{A}_{e}^{\prime}}\}\right).
\end{equation}
Considering the properties $\mathbb{P}\left(\bigcup_{i\in\mathcal{S}}S_i\right)\leq\sum_{i\in\mathcal{S}}\mathbb{P}\left(S_j\right)$ and $\mathbb{P}\left(\bigcap_{i\in\mathcal{S}}S_{i}\right)\leq\mathbb{P}\left(S_{j}\right),j\in\mathcal{S}$\cite{RIS-URA}, we have
\begin{equation}
P_{K_{s},K_{c}}\leq|\mathcal{A}_{A}|\mathbb{P}\left(\zeta_{\boldsymbol{A}_{e},\boldsymbol{A}_{a}}\right).
\end{equation}
The detection function $f_p\left(\boldsymbol{A}_d\right)$ is defined as
\begin{equation}
f_p\left(\boldsymbol{A}_d\right)=\boldsymbol{I}_L-\boldsymbol{A}_d^\mathrm{H}(\boldsymbol{A}_d\boldsymbol{A}_d^\mathrm{H})^{-1}\boldsymbol{A}_d,
\end{equation} 
manifesting the features: i) $\boldsymbol{A}_{a}f_{p}\left(\boldsymbol{A}_{a}\right)=\boldsymbol{0}_{(|\mathcal{A}_{c}|+|\mathcal{A}_{s}|),L}$, ii) $f_{p}\left(\boldsymbol{A}_{a}\right)\preceq\boldsymbol{I}_{L}$, iii) $f_p\left(\boldsymbol{A}_a\right)f_p\left(\boldsymbol{A}_a\right)^{\mathrm{H}}=f_{p}\left(\boldsymbol{A}_{a}\right)$, by which one can obtain
\begin{equation}\label{ineq}
\|\boldsymbol{Y}f_p\left(\boldsymbol{A}_a\right)\|^2\leq\|\boldsymbol{Z}\|^2.
\end{equation}
Recalling the signal model $\boldsymbol{Y}=\boldsymbol{G}_a\boldsymbol{A}_a+\boldsymbol{Z}$, it can be written into a format separating erroneous signals, i.e.,
\begin{equation}
\boldsymbol{Y}=\begin{bmatrix}\boldsymbol{G}_{a,1},\boldsymbol{0}_{M,(K_s+K_c)}\end{bmatrix}\boldsymbol{A}_e+\boldsymbol{Z}^{\prime}+\boldsymbol{Z},
\end{equation}
where $\boldsymbol{G}_{a,1}$ is a sub-matrix with codewords to be correctly detected, $\boldsymbol{Z}^{\prime}=\boldsymbol{G}_{a,2}\boldsymbol{A}_{a,2}$. Assuming independent and identically distributed (i.i.d.) array response elements, $\boldsymbol{Z}^{\prime}$ follows the distribution of $\mathcal{CN}(0,\sigma^2_t)$, where $\sigma_t^2=\frac{1}{M}\mathbb{E}\left\{\boldsymbol{g}_j\right\}K_cP_c'+L_sP_s'$. Also, from the law of large numbers, we have
\begin{equation}\label{ii}
\boldsymbol{Z}^{\prime}\boldsymbol{A}_e^\mathrm{H}\approx 0_{M\times(K_c+K_s)}, \boldsymbol{Z}f_p\left(\boldsymbol{A}_e\right)\approx\boldsymbol{Z},
\end{equation}
along with the property $\boldsymbol{A}_{a}f_{p}\left(\boldsymbol{A}_{a}\right)=\boldsymbol{0}_{(|\mathcal{A}_{c}|+|\mathcal{A}_{s}|),L}$. Hence, the left side of \eqref{ineq} can be approximated as
\begin{equation}\label{left}
\|\boldsymbol{Y}f_p\left(\boldsymbol{A}_e\right)\|^2\approx\|\boldsymbol{Z}+\boldsymbol{Z}^{\prime}\|^2.
\end{equation}
Substituting \eqref{ineq} into \eqref{left} and considering the Frobenius norm of a given matrix remains identical after matrix vectorization, we have
\begin{subequations}
\begin{align}
\mathbb{P}\left(\zeta_{\boldsymbol{A}_e,\boldsymbol{A}_a}\right)&\leq\mathbb{P}\left(\|\boldsymbol{z}+\boldsymbol{z}^{\prime}\|^2<\|\boldsymbol{z}\|^2\right)\label{error_pro_1},\\
		&\leq\mathbb{E}\left\{e^{-\lambda_{1}\|\boldsymbol{z}+\boldsymbol{z}^{\prime}\|^{2}+\lambda_{1}\|\boldsymbol{z}\|^{2}}\right\}
		\label{error_pro_2},\\
		&=\mathbb{E}\left\{\frac{e^{\lambda_1\|\boldsymbol{z}\|^2}e^{\frac{-\lambda_1\|\boldsymbol{z}\|^2}{(1+\lambda_1\sigma_t^2)}}}{(1+\lambda_1\sigma_t^2)^{LM}}\right\}
		\label{error_pro_3},\\
		&=\frac{1}{\left(1+\lambda_1\sigma_t^2-\lambda_1^2\sigma^2\sigma_t^2\right)^{LM}}
		\label{error_pro_4},\\
		&\xrightarrow[\text{Maximization}]{\text{Denominator}}e^{-LM\log\left(1+0.25\sigma_t^2/\sigma^2\right)}
		\label{error_pro_5},
	\end{align}
\end{subequations}
where $\boldsymbol{z}$ and $\boldsymbol{z}^{\prime}$ are the vectorized matrix of $\boldsymbol{Z}$ and $\boldsymbol{Z}^{\prime}$, and the following results are used:
\begin{itemize}
\item From \eqref{error_pro_1} to \eqref{error_pro_2}, the Chernoff bound is utilized, i.e., $\mathbb{P}(x>0)\leq\mathbb{E}(e^{\lambda_1x}),\lambda_1>0$.
\item From \eqref{error_pro_2} to \eqref{error_pro_3}, the following identity is used:
\begin{equation*}
\mathbb{E}\left\{e^{x\|\boldsymbol{a}+\boldsymbol{b}\|^{2}}\right\}=\frac{1}{(1-x\sigma_{a}^{2})^{L}}e^ { \frac{x\|\boldsymbol{b}\|^{2} }{ 1-x\sigma_{a}^{2} }  },
\end{equation*}
and $x\sigma_a^2<1,\boldsymbol{a}\sim\mathcal{CN}\left(\boldsymbol{0},\sigma_a^2\boldsymbol{I}_n\right)$.
\item The denominator in \eqref{error_pro_4} is a quadratic equation of one unknown $\lambda_1$ with minimum function value at $\lambda_1=\frac{1}{2\sigma^2}$.
\item Expression \eqref{error_pro_5} is the $\log$-format of \eqref{error_pro_4}.
\end{itemize}
Since $\sigma_t^2=\frac{1}{M}\mathbb{E}\left\{\boldsymbol{g}_j\right\}K_cP_c'+L_sP_s'$, the error probability $\mathbb{P}\left(\zeta_{\boldsymbol{A}_e,\boldsymbol{A}_a}\right)$ is inversely proportional to $\sigma_t^2$.

\section{$k$-Sparse Vector Estimation Under Lasso Model}\label{app:2}
Here, we derive the estimation upper bound of Lasso solutions to the $k$-sparse vector estimation. Recalling the Lasso estimation model for the sparse linear regression problem $\boldsymbol{v}=\boldsymbol{A}\boldsymbol{\beta}+\boldsymbol{n}_z$, we have
\begin{equation}
\hat{\boldsymbol{\beta}}=\underset{\|\beta\|_1\leq \varepsilon_1}{\operatorname*{\operatorname*{min}}}\|\boldsymbol{v}-\boldsymbol{A}\boldsymbol{\beta}\|_2^2,
\end{equation}
where the sensing codebook is $\boldsymbol{A}\in \mathbb{C}^{M^2\times N}$ and $\boldsymbol{\beta}\in \mathbb{C}^{M^2\times 1}$ is a $K$-sparse vector to be estimated, i.e., there will be only $k$ nonzero elements. Setups on $M,N,k$ should at least satisfy the following constraints: 1) For a sparse linear regression by compressive sensing, $N\gg K$ and $M^2\ge K$; 2) The setting of $N$ should render a codebook to satisfy the potential estimation resolution. The estimation on $\boldsymbol{\beta}$ follows the bound of 
\begin{equation}\label{bound}
\|\hat{\boldsymbol{\beta}}-\boldsymbol{\beta}\|_2\leq\frac{4}{\gamma}\sqrt{\frac{k}{M^2}}\left\|\frac{\boldsymbol{A}^\mathrm{H}\boldsymbol{n}_z}{\sqrt{M^2}}\right\|_\infty,
\end{equation}
where $\gamma$ is the eigenvalue of $\nabla^2f(\boldsymbol{\beta})=\boldsymbol{A}^\mathrm{H}\boldsymbol{A}/M^2$ to be selected to construct strong-convexity. Specifically, the objective function $\begin{array}{c}{f_{M^2}(\boldsymbol{\beta})=\frac{1}{2M^2}\|\boldsymbol{v}-\boldsymbol{A}\boldsymbol{\beta}\|_{2}^{2}}\end{array}$ will always be convex. However, when the covariance matrix $\boldsymbol{A}^{\mathrm{H}}\boldsymbol{A}$ becomes \textit{rank-deficient}, i.e., $N>M^2$, the objective function $f_{M^2}(\boldsymbol{\beta})$ is not \textit{strong-convex}, i.e., cannot guarantee a unique minimum point. 

However, from \cite[Chapter 11.2.2]{Lasso_Upper}, \textit{a convex loss function in high-dimensional settings (with $N\gg M^2$) cannot be strongly convex; rather, it will be curved in some directions but flat in others.} It also reveals that the objective function satisfies the restricted strong convexity at $\hat{\boldsymbol{\beta}}$ with respect to $\mathcal{C}(\mathcal{S},3)$ if there is a constant $\gamma >0$ such that
\begin{equation}\label{constraint}
\frac{\frac{1}{M^2}\nu\boldsymbol{A}^\mathrm{H}\boldsymbol{A}\nu}{\|\nu\|_2^2}\geq\gamma, \nu \in \mathcal{C},
\end{equation}
where $\mathcal{C}(\mathcal{S};\alpha):=\{\nu\in\mathbb{C}^k\mid\|\nu_{\mathcal{S}^c}\|_1\leq\alpha\|\nu_\mathcal{S}\|_1\}$, set $\mathcal{S}$ contains possible values of $\boldsymbol{\beta}$ and $\alpha\le 1$. 

Here, we explain how to determine an appropriate selection of $\gamma$. Considering the estimation case in this work where $\boldsymbol{\beta}$ is a $1$-sparse vector, i.e., $\mathcal{S}=\{0,1\}$ and the estimation takes place after the codeword detection, i.e., the detection error has been considered by $P_{K_s,K_c}$. Therefore, the estimation problem can be approximated and simplified into $\boldsymbol{v}=\tilde{\boldsymbol{g}}_{\theta_n}\boldsymbol{\beta}+\boldsymbol{n}_z$ where $\tilde{\boldsymbol{g}}_{\theta_n}^\mathrm{H}\tilde{\boldsymbol{g}}_{\theta_n}=M^2$, i.e., \eqref{constraint} can be calculated as $1 \geq\gamma\ge 0$.
Meanwhile, the bound at the right side of \eqref{bound} is inversely proportional to $\gamma$. Therefore, fixing $\gamma$ to 1 can get a tight estimation upper bound, i.e., the estimation bound of Lasso model can be written in \eqref{Lasso-Error-upper} as $\|\boldsymbol{\beta}-\hat{\boldsymbol{\beta}}\|_2\le \frac{4}{M}\|\frac{\boldsymbol{A}^{\mathrm{H}}\boldsymbol{n}_z}{M}\|_{\infty}$.



%

\vfill
\end{document}